\documentclass[a4paper,UKenglish]{lipics-v2016}
\usepackage{xspace}
\usepackage{amssymb}
\usepackage{amsfonts}
\usepackage{amsmath}

\newtheorem{observation}{Observation}
\renewcommand{\mod}{{\rm ~mod~}}

\usepackage{mathtools}
\DeclarePairedDelimiter{\floor}{\lfloor}{\rfloor}
\newcommand{\cc}[1]	{\textsf{#1}\xspace}
\newcommand{\YES}	{\textsc{Yes}\xspace}
\newcommand{\NO}		{\textsc{No}\xspace}

\newcommand{\problemdef}[3]{
    \begin{center}
    \noindent
    \framebox{
        \begin{minipage}{4.6in}
            \textbf{\sc #1} \\
            \emph{Input}: #2 \\
            \emph{Question}: #3
        \end{minipage}
    }
    \end{center}
}

\newcommand{\hds}{{\sc Hypergraph Degree Sequence}\xspace}
\newcommand{\NPC}{{\sf NP}-Complete\xspace}

\title{Cyclic Hypergraph Degree Sequences}

\author{S M Meesum}
\affil{The Institute of Mathematical Sciences,\\ HBNI, Chennai, India.\\
\texttt{meesum@imsc.res.in}} 


\EventEditors{John Q. Open and Joan R. Acces}
\EventNoEds{2}
\EventLongTitle{42nd Conference on Very Important Topics (CVIT 2016)}
\EventShortTitle{CVIT 2016}
\EventAcronym{CVIT}
\EventYear{2016}
\EventDate{December 24--27, 2016}
\EventLocation{Little Whinging, United Kingdom}
\EventLogo{}
\SeriesVolume{42}
\ArticleNo{23}

\begin{document}
\maketitle
\begin{abstract}
The problem of efficiently characterizing degree sequences of simple hypergraphs is a fundamental long-standing open problem in Graph Theory.
Several results are known for restricted versions of this problem. This paper adds to the list of sufficient conditions for a degree sequence to be {\em hypergraphic}.
This paper proves a combinatorial lemma about cyclically permuting the columns of a binary table with length $n$ binary sequences as rows. We prove that for any set of cyclic permutations acting on its columns, the resulting table has all of its $2^n$ rows distinct. Using this property, we first define a subset {\em cyclic hyper degrees} of hypergraphic sequences and show that they admit a polynomial time recognition algorithm. Next, we prove that there are at least $2^{\frac{(n-1)(n-2)}{2}}$ {\em cyclic hyper degrees}, which also serves as a lower bound on the number of {\em hypergraphic} sequences. The {\em cyclic hyper degrees} also enjoy a structural characterization, they are the integral points contained in the union of some $n$-dimensional rectangles.
\end{abstract}

\section{Introduction}

For a given list of positive integers $D=(d_1,\dots,d_n)\in \mathbb{Z}_+^n$, the {\em graph realization} problem asks if there exists a simple graph\footnote{A loopless graph without repeated edges.} $G_D$ on $n$ vertices whose vertex degrees are given by the list $D$.
If such a graph exists then the degree sequence $D$ is said to be {\em realizable}. 
A hypergraph $H$ is said to be a simple $k$-hypergraph if every edge has $k$ vertices. A $k$-hypergraph is called as a simple hypergraph if none of its edges are repeated and no edge contains a repeated vertex.
The case of a simple graph can be seen to correspond to the case of  $2$-hypergraphs. Generally, given a degree sequence and a positive integer $k$, one wants to find a $k$-hypergraph  {\em realizing} it.
When $k=2$, Erd\"os-Gallai Theorem \cite{gallai1960graphs} gives necessary and sufficient conditions that must be satisfied by $D$ to be {\em realizable}. 
Other characterizations for the {\em realizability} of graphs have been given by Havel \cite{havel1955remark} and Hakimi \cite{hakami1962realizability}. Sierksma and Hoogeveen  \cite{sierksma1991seven} collected seven different characterizations for graph realizability and proved their equivalence.

An input degree sequence is assumed to be given in binary encoding, and an efficient characterization requires conditions which can be checked in time which is a polynomial in input size.
For $k\geq3$, the problem becomes difficult in the sense that there is no known efficient characterization of {\em realizable} degree sequences. A characterization was given by Dewdeny \cite{dewdney1975degree} for all $k\geq 3$, which does not yield an efficient algorithm. Recently, some sharp sufficient conditions for {\em realizability} of a degree sequence based on a sequence's length and degree sum were given \cite{behrens2013new}. For a fixed value of $k\geq 3$, the problem is easily seen to be in {\sf NP}, but neither a polynomial time algorithm nor a hardness proof is known for this class of problems.
However, Colbourn, Kocay and Stinson \cite{colbourn1986some} proved that several other problems related to $3$-graphic sequences are \NPC.
Achuthan et al \cite{achuthan19933}, Billington \cite{billington1988conditions} and Choudum \cite{choudum1991graphic} gave several necessary conditions for $3$-hypergraphs, however Achuthan et al \cite{achuthan19933} also showed that none of these conditions are sufficient.
There are many surveys available for this problem \cite{hakimi1978graphs,rao1981survey,tyshkevich1987graphs1,tyshkevich1988graphs2,tyshkevich1988graphs3},
for a recent survey on related problems see \cite{ferrara2013some}.

If we give up the restriction on sizes of edges to be $k$, we get the \hds problem which was first stated in~\cite{berge1984hypergraphs} for a restricted class of hypergraphs. In particular, the \hds problem, given a list of degrees $D$, asks if there exists a simple hypergraph which is a {\em realization} of $D$. This problem appears to be harder than the class of $k$-hypergraph problems in the sense that it not even known to be in {\sf NP}.
It can be easily seen to be in \cc{PSPACE}.
Several restricted versions of this problem have been studied in the past.
Bhave, Bam, Deshpande~\cite{bhave200813}
gave an Erd\"os-Gallai-type characterization of degree sequences of loopless linear hypergraphs, where a linear hypergraph is one in which any two edges have at most one common vertex.
In another direction,
characterization for a partial Steiner triple system (PSTS), which is a linear 3-hypergraph, were given by Keranen et al~\cite{keranen2009degree}. The results in \cite{bhave200813} and~\cite{keranen2009degree} were recently generalized by Khan~\cite{} using partial $(n,k,\lambda)$-systems.

This paper provides a sufficient condition for a degree sequence to {\em realizable} by a hypergraph. We do not place any restriction on hypergraph being a linear hypergraph or a have bounded edge intersections like the cases studied earlier.

\begin{table}
\caption{$n$-Bit-Table for $n=3$}
\centering
\begin{tabular}{|c||c|c|c|}
\hline
$\times$ & $c_{3,3}$ & $c_{2,3}$ & $c_{1,3}$\\
\hline
\noalign{\smallskip}
\hline
0 & 0 & 0 & 0\\
1& 0 & 0 & 1\\
2& 0 & 1 & 0\\
3& 0  & 1 & 1\\
4& 1 & 0 & 0\\
5& 1 & 0 & 1\\
6& 1 & 1 & 0\\
7& 1 & 1 & 1\\
\hline
\end{tabular}
\label{table:bitlist3}
\end{table}

\paragraph*{Our results and Contribution}

We define a notion of cyclic permutations of a binary table (for example see Table~\ref{table:bitlist3})
and use it to find an efficiently computable sufficient condition for a given degree sequence to be {\em realizable} by a simple hypergraph.

\begin{enumerate}
\item
Firstly, we prove that for any set of cyclic permutations acting on the columns of a binary table, the resulting table has all of its $2^n$ rows distinct. 
\item
 Next, we define a notion of {\em cyclic hyper degrees} viz. the degree sequences which are the sum of contiguous rows in a  binary table in which each column has been permuted by different cyclic permutations. As cyclic permutations of binary tables have every row distinct, a {\em cyclic hyper degree} sequence is {\em realizable} for some simple hypergraph. Then we give an efficient algorithm which checks if a given degree sequence is a {\em cyclic hyper degree}. An equivalent definition of {\em cyclic hyper degrees} in Theorem \ref{thm:chd} shows their structure, the {\em cyclic hyper degrees} are contained in the union of some $n$-dimensional rectangles.
\item
Finally, we provide a lower bound of $2^{\frac{(n-1)(n-2)}{2}}$ for the number of {\em cyclic hyper degree} sequences, by proving the existence of a large $n$-dimensional rectangle such that every integral tuple contained in it is a realizable degree sequence. This also gives us a lower bound on the number of {\em realizable} sequences. For comparision, there is an upper bound of $\mathcal{O}(2^{n\cdot(n-1)})$ for the number of {\em realizable} hypergraph degree sequences. 
\end{enumerate}
\section{Preliminaries}
The set of non-negative integers is denoted using the symbol $\mathbb{Z}_+$.
The set of integers $\{1,\dots,n\}$ is denoted by $[n]$. For $m, M\in \mathbb{Z}_+$, we refer to an integral interval $\{i : m\leq i\leq M\}$ as a range and specify it by providing its minimum and maximum element.
An $n$-tuple, also simply referred to as a list, is $L=(\ell_1,\ell_2,\dots,\ell_n)$, of size $n$, is an ordered collection of elements. We refer to its $i^{th}$ element $\ell_i$ as $L(i)$. We index any list or tuple with natural numbers starting with $1$.
The notation $a_{\times m}$ denotes the tuple
$(a,\dots,a)$ consisting of $a$ repeated $m$ times.
We use $L_1 \cdot L_2$ to denote the list obtained by the operation of concatenating two lists $L_1$ and $L_2$.
A table $T=[L_1,L_2,\dots,L_m]$ is a collection of lists of equal size. Pictorially, the lists $L_1,\dots,L_m$ are arranged as columns in the table $T$. If each list $L_i$ is of size $n$, then the table $T$ is said to be of size $n\times m$. We refer to the $i^{th}$ row of a table $T$ using the notation $T(i)$ and to an $(i,j)^{th}$ entry using the symbol $T(i,j)$.

For ease of notation, the sum of the entries in a list $L$ will be denoted by $\sum L = \sum_{i\in [\vert L\vert]} L(i)$.
The sum of $L=(\ell_1,\dots,\ell_n)$ and $L'=(\ell'_1,\dots,\ell'_n)$, denoted by $L+L'$, is the $n$-tuple $(\ell_1+\ell'_1,\dots,\ell_n+\ell'_n)$. If $S$ is a list of lists then the sum of its elements will be denoted using $\sum S = \sum_{x\in S} x$.


For $k\geq 0$, a permutation $\pi$ is called a cyclic permutation of order $k$, for each $i\in [n]$ it maps $i \mapsto 1+((i+k-1)\mbox{ mod }n)$.

A hypergraph $H$ is a pair $([n],\mathcal{F})$, where $\mathcal{F}$ is a family of subsets of $[n]$. A hypergraph is simple if no set is repeated in $\mathcal{F}$. The degree of a vertex $v\in[n]$ is equal to $\vert\{F\in\mathcal{F}~:~v\in F\}\vert$.
We will be working with an equivalent version, which can be stated in terms of co-ordinate wise sum of binary sequences of length $n$.

For a given positive integer $n$, consider the set $S_n = \{0,1\}^n$ consisting of all binary tuples of length $n$. The elements of $S_n$ will also be referred to as binary sequences.
We construct a set $H_n = \{ \sum_{x \in S} x : \varnothing \subseteq S\subseteq S_n\}$. Note that for the empty set $\varnothing$ the corresponding sum $\sum_{x\in\varnothing} x$ is equal to $0_{\times n}$. By construction, each element of $H_n$ is {\em realized} by some simple hypergraph and the degree sequence of every simple hypergraph is contained in $H_n$. Each element of $H_n$ is said to be {\em representable} or is said to admit a {\em representation}.
Given this setting the {\em realizability} problem for simple hypergraphs can be restated as follows.

\problemdef
{\hds}
{A tuple $w \in \mathbb{Z}_+^n$ which is provided as a binary input.}
{Is $w \in H_n$ ?}

If $w=(w_1,\dots,w_n) \in H_n$ and $w_n=0$, then $(w_1,\dots, w_{n-1})\in H_{n-1}$. As $\sum S_n = 2^{n-1}_{\times n}$, the maximum possible value of any entry in $w$ is $2^{n-1}$. Thus, if any entry of $w$ is outside the range $\{0, \ldots, 2^{n-1}\}$, then it is not a member of $H_n$.
Even though the number of subsets of $S_n$ is $2^{2^n}$, we get that the cardinality of $H_n$ is at most $2^{n\cdot (n-1)}$.

\section{Cyclic permutations and Binary Tables}
In this section, we will be working with binary tables of size $2^n \times n$ and study the action of cyclic permutations on the columns of the table. For a given number $n \in \mathbb{Z}_+$, list out the binary expansion of numbers in increasing order from $\{0,\dots,2^n-1\}$ as rows in a table. We pad the binary expansion with sufficient numbers of zeros on the left to make the length of each row exactly equal to $n$.
For example, when $n=3$, the table is as given in Table~\ref{table:bitlist3}.

To state it formally we need the following definitions. Given a number $m$ we denote the $i^{th}$  bit in its binary representation by $\rm{bin}(m,i)$. If the most significant bit in the binary expansion occurs at the $s^{th}$-position in the binary expansion of $m$, then for all values of $i>s$ the value of $\rm{bin}(m,i)$ is zero. For example, $\rm{bin}(4,2)=1$ and $\rm{bin}(4,i)=0$ for every $i\geq 3$.

For a given $n$, we construct $n$ lists $c_{1,n}, \dots, c_{n,n}$, with each list $c_{i,n}$ having length equal to $2^n$. 
For $n=2$, we have $c_{1,2} = (0,1,0,1)$ and $c_{2,2} = (0,0,1,1)$.
For $n=3$, the lists $c_{1,3},c_{2,3},c_{3,3}$ correspond to the columns of the Table~\ref{table:bitlist3}. 
Formally, the lists are defined as follows.

\begin{definition}[$n$-Bit-Lists]
 For a given $n$, we define $n$-Bit-List to consist of $n$ lists $c_{1,n}, \dots, c_{n,n}$. For $j\in [2^n]$, the value of $c_{i,n}(j)$ is equal to $\rm{bin}(j-1,i)$.
\end{definition}

\begin{definition}[$n$-Bit-Table]
 For a positive integer $n$, the $n$-Bit-Table $T_n$ is defined to be a size $2^n\times n$ table with $T_n=[c_{n,n},c_{n-1,n},\dots,c_{1,n}]$. 
\end{definition}

The lists $c_{1,n}, \dots, c_{n,n}$ have a nice recursive structure and can be generated in an alternative way by concatenation.
Given a positive integer $n$, the base case of $n=1$ is one list $c_{1,1}=(0,1)$.
For $n\geq 2$, the tuple $c_{n,n}=0_{\times 2^{n-1}}\cdot 1_{\times 2^{n-1}}$ and for $j\in [n-1]$, the list $c_{j,n}$ is equal to the concatenated list $c_{j,n-1}\cdot c_{j,n-1}$. Thus, we get the following observation about the lists.

\begin{observation}
 \label{obs:zero-one-power}
 For $n\in\mathbb{Z}_+$ and $i\in [n]$, we have
  $c_{i,n}= (0_{\times 2^{i-1}} \cdot 1_{\times 2^{i-1}})_{\times 2^{n-i}}$.
\end{observation}

Let $\mathcal{C}_n$ be the set of all cyclic permutations on an $2^n$ length list.
Let $\Pi \subseteq \mathcal{C}_n$ be a multi-set consisting of $n$ arbitrary cyclic permutations $\pi_1, \ldots, \pi_n$.
Let $\Pi(T_n)=[\pi_n(c_{n,n}),\dots,\pi_1(c_{1,n})]$ be the new table obtained from $T_n$. 
For clarity we will change the notation slightly, let $\Pi(T_n,i)$ denote row $i$ of $\Pi(T_n)$, it is a length $n$ binary tuple
$(c_{n,n}(\pi_n(i)),  c_{n-1,n}(\pi_{n-1}(i)), \ldots, c_{1,n}(\pi_1(i)))$,
for $i \in [2^n]$.

We next state a general lemma whose proof follows by induction over the order of a cyclic permutation.

\begin{lemma}
\label{lem:halfcycle}
Let $L$ be a list and $\pi$ be any cyclic permutation, then $\pi(L\cdot L)= \pi(L)\cdot \pi(L).$
\end{lemma}

Recall that for $j\in [n-1]$, the tuple $c_{j,n}$ consists of two copies of $c_{j,n-1}$ concatenated together. Combining this fact with Lemma~\ref{lem:halfcycle}, we obtain the following as a special case.

\begin{corollary}
\label{cor:halfcycle}
 Given a positive integer $n$ and a cyclic permutation $\pi \in \mathcal{C}_n$.
 For $j \in [n-1]$, the tuple $\pi(c_{j,n})$ is equal to $\pi(c_{j,n-1}) \cdot \pi(c_{j,n-1})$.
\end{corollary}

We next prove that for any set of cyclic permutations $\Pi$, any two rows in $\Pi(T_n)$ will never become equal.

\begin{theorem}\label{thm:rotatebits}
For any list of $n$ cyclic permutations $\Pi$, the rows of $\Pi(T_n)$ are pair-wise distinct.
\end{theorem}
\begin{proof}
 We prove this by induction on $n$. For the base case $n=1$, the statement is  trivially true. For the rest of the proof, assume that the list of cyclic permutations is $\Pi=(\pi_1,\ldots,\pi_n)$ and $\Pi_{\overline{n}}$ is used to denote the list $(\pi_1,\dots,\pi_{n-1})$.
 
The table $T_n$ can be constructed recursively by taking two copies of $T_{n-1}$ and appending the rows of one below the other, after that we add $0_{\times 2^{n-1}}\cdot 1_{\times 2^{n-1}}$ as the first column. Consider the table 
\begin{align*}
T_{\overline n} &= [c_{n-1,n},\dots,c_{1,n}] \\
             &= [c_{n-1,n-1}\cdot c_{n-1,n-1},\dots, c_{1,n-1}\cdot c_{1,n-1}].
\end{align*}
 Apply the list of permutations $\Pi_{\overline{n}}$ on $T_{\overline n}$ to get
\begin{align*}
 \Pi_{\overline{n}}(T_{\overline n}) &= [\pi_{n-1}(c_{n-1,n}), \ldots, \pi_1(c_{1,n})]\\
 			&= [\pi_{n-1}(c_{n-1,n-1})\cdot \pi_{n-1}(c_{n-1,n-1}),\dots, \pi_1(c_{1,n-1})\cdot \pi_1(c_{1,n-1})],
\end{align*}
 where the second equality follows from Corollary~\ref{cor:halfcycle}.
 By the induction hypothesis, the first row-wise half of $\Pi_{\overline{n}}(T_{\overline n})$, which is the same as $\Pi_{\overline{n}}(T_{n-1})$, consists of distinct rows. 
 Therefore, the table $\Pi_{\overline{n}}(T_{\overline n})$ consists of rows which are repeated exactly twice. For $i < j$, the rows $\Pi_{\overline{n}}(T_{\overline n},i)$ and
 $\Pi_{\overline{n}}(T_{\overline n},j)$ are equal when $j=i+2^{n-1}$.
Therefore, it suffices to prove that the rows  $\Pi(T_n,i)$ and $\Pi(T_n,i+2^{n-1})$ are distinct.
 Observe that to obtain the table $\Pi(T_n)$, we need to append $\pi_n(c_{n,n})$ as the first column in $\Pi_{\overline{n}}(T_{\overline n})$. As $c_{n,n}$ is equal to $0_{\times 2^{n-1}}\cdot 1_{\times 2^{n-1}}$, we have $c_{n,n}(i)\neq c_{n,n}(i+2^{n-1})$, this implies that $c_{n,n}(\pi_n(i))\neq c_{n,n}(\pi_n(i+2^{n-1}))$.
\end{proof}

Next, we define a notion of {\em cyclic hyper degrees} as follows.

\begin{definition}[Cyclic Hyper Degree]\label{def:chd}
 Given $\Pi$, a list of $n$ cyclic permutations, a tuple $d \in \mathbb{Z}_+^n$ is said to be a {\em cyclic hyper degree} if there exist $i, N \in [2^n]$ such that $d=\sum_{k=i}^N \Pi(T_n, k).$
\end{definition}

As the rows of $\Pi(T_n)$ are distinct, their contiguous sum  is in $H_n$ by definition. This gives us the main theorem of this section.

\begin{theorem}
 If $w\in \mathbb{Z}_+^n$ is a {\em cyclic hyper degree}, then $w\in H_n$.
\end{theorem}

We note that $(4,1,1,1)$ is a {\em realizable} hypergraph degree sequence but it is not a {\em cyclic hyper degree} sequence.
In the next section, we will show how to efficiently check if a given sequence $d$ is a {\em cyclic hyper degree}. 

\section{Efficiently Recognizing {\em Cyclic Hyper Degrees}}
This section consists of two parts. In the first part will culminate with Theorem~\ref{thm:shiftrange} which provides an efficiently computable closed form formula for the range of values taken by contiguous sum of $N$ elements in a list $c_{i,n}$, for any $i$. In the second part we will show how to use Theorem~\ref{thm:shiftrange} to decide if a given degree sequence is a {\em cyclic hyper degree}.

The elements in the columns of $T_n$ do not change their relative position after application of a cyclic permutation when seen as a cyclic list. We shall use this property to efficiently search for possible bit subsets which may sum up to a given input degree sequence.

\subsection{Contiguous Sum of Bit Lists}

\begin{definition}[Contiguous Sum]
\label{def:csum}
 Given a list $L$ of length $m$, the contiguous sum of $N$ elements in $L$ starting at the index $i\in [m]$ is defined to be
 $$\mathcal{S}(L,i,N):=\sum_{j=0}^{N-1} L(1+ ((i+j-1)\mod m)).$$
\end{definition}

The summation above treats the list $L$ as a cyclic list. Next, we prove that the contiguous sum function is a `continuous' function, this property will allow us to specify the range of sum by stating the minimum and the maximum value taken by it.
 Note that if $L$ is a $0$-$1$ list, for any index $\ell \in [m]$, 
 we have $\vert \mathcal{S}(L,\ell,N)-\mathcal{S}(L,\ell+1,N) \vert \in \{0,1\}$. This fact gives us the following property.

\begin{observation}
\label{obs:continuity}
 Let $L$ be a size $m$ list having $0$-$1$ entries and $N\in \mathbb{Z}_+$.
 If $v_i=\mathcal{S}(L,i,N)$ and $v_j=\mathcal{S}(L,j,N)$, for some $i,j\in[m]$, then 
 for every $v\in \mathbb{Z}_+$ contained between  $v_i$ and $v_j$
 there exists a $k\in [m]$  such that $\mathcal{S}(L,k,N)=v$.
\end{observation}

As the lists $c_{i,n}$ are over $0$-$1$ we get an easy relation between the maximum and minimum values taken by the contiguous sum as follows.

\begin{lemma}
\label{lem:sumcomplement}
  Let  $j\in\{0,\dots,n\}$, $i \in [n]$ and $N\in[2^n]$. The minimum of the sum of $N$ contiguous bits in a bit list $c_{i,n}$ is $m$ if and only if its maximum is $N-m$.
\end{lemma}
\begin{proof}
 Let $\overline{c}_{i,n}$ be the bit list obtained from the list $c_{i,n}$ by flipping each zero to one and vice versa. Let $\sigma_{2^{i-1}}$ be an order $2^{i-1}$ cyclic permutation, observe that $\overline{c}_{i,n}$ is equal to $\sigma_{2^{i-1}}(c_{i,n})$.
 If the minimum value is obtained at the contiguous segment which starts at the index $j$ in $c_{i,n}$, then the value $N-m$ can be obtained by the contiguous sum starting at index $\sigma_{2^{i-1}}(j)$. Finally, note that $m$ is the minimum value if and only if $N-m$ is the maximum value.
\end{proof}

Combining Observation~\ref{obs:continuity} and Lemma~\ref{lem:sumcomplement} we get the following.

\begin{lemma}
\label{lem:minmax}
 Let $N\in[2^n]$ and $m=\min_{j \in [2^n]} \mathcal{S}(c_{i,n}, j, N)$. For every value $v$ in the range $\{m,\dots, N-m\}$ there exists a $j\in [2^n]$ such that $\mathcal{S}(c_{i,n}, j, N)=v$.
\end{lemma}

The lemma above allows us to find the range of values taken by the contiguous sum by just finding the minimum value taken by it.
Next we prove a simpler lemma about the range of values taken. Using that, in Theorem~\ref{thm:shiftrange}, we will find the range of values taken by the contiguous sum of $N$ elements in any list $c_{i,n}$.

\begin{lemma}
\label{lem:shiftpower}
 For  $j\in\{0,\dots,n\}$ and $i \in [n]$, the sum of $2^j$ contiguous bits in a bit list $c_{i,n}$ takes the following values.
\begin{enumerate}
\item
\label{enum:shiftpower-one}
If $j \leq (i-1)$, then the range is $\{0, \dots,2^j \}$, and
\item
\label{enum:shiftpower-two}
If $j\geq i$, then the sum is exactly $2^{j-1}$.
\end{enumerate}
\end{lemma}
\begin{proof}
 By Lemma~\ref{lem:minmax}, it suffices to find the minimum value of contiguous sum function. Notice that we have, $c_{i,n}= (0_{\times 2^{i-1}} \cdot 1_{\times 2^{i-1}})_{\times 2^{n-i}}$, by Observation~\ref{obs:zero-one-power}.
 \begin{enumerate}
 \item
 When $j\leq (i-1)$, we can pick a block of $2^j$ zeros giving a total of zero, which is the minimum possible value.
 \item
 When $j\geq i$, let $L_k$ be a list of $2^j$ contiguous bits of $c_{i,n}$ starting at the index $k$ in $c_{i,n}$. To prove that $\sum L_k = \sum L_{k+1} $, it suffices to show that $c_{i,n}(k)=c_{i,n}(k+2^j)$. Rewriting
 $c_{i,n}=((0_{\times 2^{i-1}}\cdot 1_{\times 2^{i-1}})_{\times 2^{j-i}})_{\times 2^{n-j}}$ shows that any two indices with difference equal to $2^j$ store the same value. As the choice of $k$ was arbitrary, the contiguous sum is equal to $2^{j-1}$.
 \end{enumerate} 
\end{proof}

\begin{theorem}
\label{thm:shiftrange}
 For $i \in [n]$, $N\in[2^{n}]$ and $p=2^i$, the sum of $N$ contiguous bits in a bit list $c_{i,n}$ takes values in the range, ${\rm range}(i,N) \triangleq$
 $$ 
 \left\{
	\floor[\Big]{\frac{N}{p}} \frac{p}{2} + \max \left( (N\mod p) - \frac{p}{2}, 0 \right),
  	\cdots,
   	\floor[\Big]{\frac{N}{p}} \frac{p}{2} + \min \left( N\mod p, \frac{p}{2} \right)
 \right\}.
 $$
\end{theorem}
\begin{proof}
 For a fixed $i\in[n]$ consider the list $c_{i,n}$. Assuming that the minimum value of the range is as claimed, by Lemma~\ref{lem:minmax}, the maximum value is
\begin{align*}
   \max_{j \in [2^n]} \mathcal{S}(c_{i,n}, j, N)
   &= N-\min_{j \in [2^n]} \mathcal{S}(c_{i,n}, j, N)\\
   &=N- \left( \floor[\Big]{\frac{N}{p}} \frac{p}{2} + \max \left( (N\mod p) - \frac{p}{2}, 0 \right)\right)\\
 &= \floor[\Big]{\frac{N}{p}}p + (N\mod p) - 
 \left( 
 \floor [\Big] {\frac{N}{p}} \frac{p}{2} + \max \left( (N\mod p) - \frac{p}{2}, 0 \right)
 \right)	\\
 &= \floor[\Big]{\frac{N}{p}} \frac{p}{2} +(N\mod p) - 
 \max	\left(		(N \mod p) - \frac{p}{2}, 	0 	\right)	\\
 &= \floor[\Big]{\frac{N}{p}} \frac{p}{2} + \min \left( (N\mod p)-(N \mod p) + \frac{p}{2}, N\mod p\right)\\
 &=  \floor[\Big]{\frac{N}{p}} \frac{p}{2} + \min \left(\frac{p}{2}, N\mod p\right).
\end{align*}
 
 As proved in case~\ref{enum:shiftpower-two} of Lemma~\ref{lem:shiftpower}, the sum of $\floor{\frac{N}{p}}p$ contiguous bits is equal to $\floor{\frac{N}{p}}\frac{p}{2}$  irrespective of the starting index.
 Therefore, it suffices to find the minimum sum of $R=(N\mod 2^i)$ contiguous bits. Let $L_k$ be a list of $R$ bits occurring contiguously in $c_{i,n}$ starting at index $k$. If the first bit of $L_k$ is $1$, then $\sum L_{k+1}\leq \sum L_k$. Therefore, we can keep on increasing the value of $k$ until the first bit is zero,  without increasing the value of the contiguous sum. On the other hand, if $c_{i,n}(k-1)=0$, then $\sum L_{k-1}\leq \sum L_k$. Therefore, we can keep on decreasing the value of $k$ one at a time until $c_{i,n}(k-1)=1$, without increasing the value of the contiguous sum. Thus the minimum value of the contiguous sum is achieved when the index $k$ points to the start of any block $0_{\times 2^{i-1}}$ contained in $c_{i,n}$. The value of minimum is  $\max(R-\frac{p}{2},0)$ as the ones start appearing after $\frac{p}{2}$ indices from the start of a list $0_{\times 2^{i-1}}\cdot 1_{\times 2^{i-1}}$.  Adding it to $\floor{\frac{N}{p}}\frac{p}{2}$ gives the required minimum value.
\end{proof}

\subsection{Algorithm}
We next state a theorem which gives an equivalent definition of {\em cyclic hyper degrees}.
\begin{theorem}
\label{thm:chd}
A list $w=\{w_1,\dots,w_n\}\in \mathbb{Z}_+^n$ is a {\em cyclic hyper degree} if and only if there exist $N\in[2^n]$ and a permutation $\pi$, such that for each $i\in [n]$, $w_{\pi(i)} \in {\rm range}(i,N)$.
\end{theorem}

\begin{proof}
 Forward direction is a direct consequence of the definition.
 
 Using Definitions~\ref{def:chd} and~\ref{def:csum}, we get that there exist numbers $s_1,\dots,s_n \in [2^n]$ such that for each $i\in [n]$, we have $w_{\pi(i)}=\mathcal{S}(c_{i,n},s_i,N)$. Let $\Pi^{-1}=(\sigma_{s_1}^{-1},\dots,\sigma_{s_n}^{-1})$ be the list of cyclic permutations, where for each $i\in [n]$, $\sigma_{s_i}^{-1}$ is the inverse of the cyclic permutation of order $s_i$. Consider the table $\Pi^{-1}(T_n)$, by
 Theorem~\ref{thm:rotatebits}, all its rows are distinct. In particular, the first $N$ rows are distinct and their sum is $\pi(w)$. Finally, $w\in H_n$ if and only if $\pi(w)\in H_n$.
\end{proof}

Theorem~\ref{thm:shiftrange} gives us a way to efficiently find the number of bits in a contiguous sum of $N$ bits. If we know the number of distinct bit sequences that can sum up to a given vector $w\in \mathbb{Z}_+^n$, then using Theorem~\ref{thm:shiftrange} we can generate all the possible ranges of values which can be taken by each coordinate of the sum. Finally, we need to check if each coordinate of $w$ is contained in different ranges, this corresponds to finding the permutation $\pi$ in Theorem~\ref{thm:chd}. In the next lemma, we will find the number of possible  distinct bit-sequences which can sum up to a given $w$ using cyclic shifts, this corresponds to finding $N$ in Theorem~\ref{thm:chd}.
\begin{lemma}
\label{lem:set-size}
If $w=\{w_1,\dots,w_n\}\in \mathbb{Z}_+^n$ is a {\em cyclic hyper degree}, then the number of bit sequences which sum up to $w$ is an element of the set
$$\mathcal{N}_w\triangleq \{2w_i+j~:~i\in[n], j\in \{-1,0,1\}\}.$$
\end{lemma}
\begin{proof}
 As one of the coordinates of $w$, say $w_k$, is the contiguous sum of $c_{1,n}$, we need to find the number of bits which sum up to $w_k$. From the structure of $c_{1,n}$, it is easily seen that there are just three values viz. $2w_k-1,2w_k,2w_k+1$ which contain $w_k$ in their range of sums. Conversely, for any number $x$ not contained in $\{2w_i+j~:~i\in[n], j\in \{-1,0,1\}\}$, the sum of $x$ contiguous bits $c_{1,n}$ will not contain any of $w_i$, for $i\in [n]$.
\end{proof}

\begin{lemma}
\label{lem:embed}
 Given $w\in\mathbb{Z}_+^n$ and a list of integer intervals $R_1,\dots,R_n \subset \mathbb{Z}_+^2$. There exists an algorithm running in time polynomial in $n$ which correctly answers if there exists a permutation $\pi$ such that  for each $i\in [n]$, $w_{\pi(i)}\in R_i$. 
\end{lemma}
\begin{proof}
 Construct a bipartite graph $G=(A,B,E)$ on $2n$ vertices. Let $A=B=[n]$ and $(i,j)\in E$ if and only if $w_i \in R_j$. Using a polynomial time algorithm one can find if there exists a perfect matching in $G$. If there is a perfect matching then the answer is \YES, otherwise it is \NO.
\end{proof}

\begin{theorem}
 \label{thm:chd-poly}
 There is a polynomial time algorithm in $n$ which decides if a given $w\in\mathbb{Z}_+^n$ is a {\em cyclic hyper degree}.
\end{theorem}

\begin{proof}
 For each $N\in\mathcal{N}_w$, given by Lemma~\ref{lem:set-size}, and $i\in[n]$ compute ${\rm range}(i,N)$ as given by Theorem~\ref{thm:shiftrange}. Now, use Lemma~\ref{lem:embed} on these ranges of numbers and decide if $w$ is a {\em cyclic hyper degree}, if it is not then try the next number from the set $\mathcal{N}_w$. If it succeeds for at least one element of $\mathcal{N}_w$, we answer \YES, otherwise we answer \NO. Finally, note that $\vert N_w \vert\leq 3n$ and all the other steps can be performed in time which is a polynomial function of $n$. 
\end{proof}
\section{Lower Bound on the number of {\em Cyclic Hyper Degrees}}

In this section we will lower bound the number of {\em cyclic hyper degrees} for a given value of $n$.
For a given $N$ we will find the size of range of contiguous sum, denoted by $R_{i,N}$, for each value of $i$ as given by Theorem~\ref{thm:shiftrange}. The number of distinct {\em cyclic degree sequences} which are the sum of $N$ bit sequences is then $\prod_{i\in [n]} R_{i,N}$. This is so because we have $R_{i,N}$ choices for the $i^{th}$ coordinate.
We proceed to find $R_{i,N}$, the size of each range, as a corollary of Theorem~\ref{thm:shiftrange} as follows.

\begin{corollary}
\label{cor:sumrange}
 For $i \in [n]$, $N\in[2^{n}]$ and $p=2^i$, the sum of $N$ contiguous bits in a bit list $c_{i,n}$ takes $1+ \min (N\mod p, p-  (N\mod p))$  distinct values.
\end{corollary}
\begin{proof}
 The number of values is  $1+\max_{j \in [2^n]} \mathcal{S}(c_{i,n}, j, N) -\min_{j \in [2^n]} \mathcal{S}(c_{i,n}, j, N)$
\begin{align*}
   &= 1+N-2\min_{j \in [2^n]} \mathcal{S}(c_{i,n}, j, N)\\
   &= 1+N-\floor[\Big]{\frac{N}{p}}p - 2\max \left( (N\mod p) - \frac{p}{2}, 0 \right)\\
   &= 1+N\mod p - 2\max \left( (N\mod p) - \frac{p}{2}, 0 \right)\\
   &= 1+ \min \left( N\mod p - 2 (N\mod p) +p, N\mod p \right)\\
   &= 1+ \min \left( p-  (N\mod p), N\mod p \right).
\end{align*}
\end{proof}

\begin{lemma}
\label{lem:lower}
 For $n\in \mathbb{Z}_+$, the number of {\em cyclic hyper degrees} is at least $2^{\frac{(n-1)(n-2)}{2}}$.
\end{lemma}
\begin{proof}
Given a fixed number $n$, we are going to count the number of {\em cyclic hyper degrees} which are the sum of exactly $M$ bit-sequences, where $M= \sum_{j=0}^{\floor{\frac{n}{2}}} 2^{2j}$. By Corollary~\ref{cor:sumrange}, the range of values possible for $i^{th}$ bit is $B_i=1+\min(2^i-M \mod 2^i, M \mod 2^i)=1+\min(2^i-\sum_{j=1}^{\floor{\frac{i}{2}}} 2^{2j},\sum_{j=1}^{\floor{\frac{i}{2}}} 2^{2j})$. Depending on whether $i$ is even or odd, it can be broken into two cases,  but in both the cases, for $i\geq 2$ we have $B_i\geq 2^{i-2}$.
The number of representable bit sequences possible with these ranges of coordinates is
$$
\prod_{k\in [n]} B_k
\geq \prod_{k=2}^n 2^{k-2}
= \prod_{k=0}^{n-2} 2^k
= 2^{\frac{(n-1)(n-2)}{2}}.
$$
\end{proof}

\section{Conclusion}

We looked at the \hds problem and gave a sufficient condition for hypergraphic sequences.
In particular, we proved that there is a subset of hypergraphic sequences having size at least $2^{\frac{(n-1)\cdot (n-2)}{2}}$ which is efficiently recongnizable.
It would be interesting to look at the hypergraphic sequences which are not {\em cyclic hyper degrees}, these sequences may be helpful in finding hardness reductions for it.
Lastly, it would be interesting to further study the structure of {\em cyclic hyper degrees}.

\bibliographystyle{plainurl}
\bibliography{ref}

\end{document}